\documentclass[aps,preprint,showpacs,amsmath,amssymb,floatfix]{revtex4-1}
\usepackage{graphicx}
\usepackage{color}
\usepackage{dcolumn}
\usepackage{xspace} 
\usepackage{hyperref}
\usepackage{epsfig}
\usepackage[T1]{fontenc}
\usepackage{soul} 

\def\Eq#1{Eq.~(\ref{#1})}
\def\Fig#1{Fig.~\ref{#1}}
\def\Ref#1{Ref.~\cite{#1}}
\def\Tab#1{Table~\ref{#1}}
\def\Sec#1{Section~\ref{#1}}

\begin{document}

\title{Photonuclear reactions with Zinc: A case for clinical linacs}
\author{I.~Boztosun$^{1,2}$}
\author{H.~{\DJ}apo$^{1,2}$}
\email{haris@akdeniz.edu.tr}
\homepage{http://nukleer.akdeniz.edu.tr/en}
\author{M.~Karako\c{c}$^{1,2}$}
\author{S.F.~\"{O}zmen$^{1,2}$}
\author{Y.~\c{C}e\c{c}en$^{2,3}$}
\author{A.~\c{C}oban$^{1,2}$}
\author{T.~Caner$^{1,2}$}
\author{E.Bayram$^{4}$}
\author{T.R.~Saito$^{5,6,7}$}
\author{T.~Akdo\u{g}an$^{8}$}
\author{V.~Bozkurt$^{9}$}
\author{Y.~Ku\c{c}uk$^{1,2}$}
\author{D.~Kaya$^{1,2}$}
\author{M.N. Harakeh$^{10}$}
\affiliation{$^1$ Department of Physics, Akdeniz University, TR-07058, Antalya, Turkey}
\affiliation{$^2$ Nuclear Research and Application Center, Akdeniz University, TR-07058, Antalya, Turkey}
\affiliation{$^3$ Department of Radiation Oncology, Akdeniz University,TR-07058, Antalya, Turkey}
\affiliation{$^4$ Department of Chemistry, Akdeniz University, TR-07058, Antalya, Turkey}
\affiliation{$^5$ GSI Helmholtz Center for Heavy Ion Research, Planckstrasse 1, 64291 Darmstadt, Germany}
\affiliation{$^6$ Johannes Gutenberg-Universit\"at Mainz, J.J.Becherweg 40, 55099 Mainz}
\affiliation{$^7$ The Helmholtz Institute Mainz (HIM), J.J.Becherweg 40, 55099 Mainz, Germany}
\affiliation{$^8$ Department of Physics, Bogazici University, 34342 Istanbul, Turkey}
\affiliation{$^9$ Nigde University, 51100 Nigde, Turkey}
\affiliation{$^{10}$ KVI-CART, University of Groningen, 9747 AA Groningen, Netherlands}


\date{\today}

\begin{abstract}
The use of bremsstrahlung photons produced by a linac to 
induce photonuclear reactions is wide spread. However, using a 
clinical linac to produce the photons is a new concept.
We aimed to induce photonuclear reactions on zinc isotopes 
and measure the subsequent transition energies and half-lives. 
For this purpose, a 
bremsstrahlung photon beam of 18 MeV endpoint energy produced by the
Philips SLI-25 linac has been used. The
subsequent decay has been measured with a well-shielded single HPGe
detector. The results obtained for transition energies are in good agreement
with the literature data and in many cases surpass these in accuracy. For the
half-lives, we are in agreement with the literature data, but do not achieve their precision.
The obtained accuracy for the transition energies show what is achievable 
in an experiment such as ours. We demonstrate the usefulness and
benefits of employing clinical linacs for nuclear physics experiments.
\end{abstract}
\maketitle

%
%
%

%
\section{Introduction}
\label{sec:intro}

The study of photon-induced nuclear reactions has been 
of interest over the years. The motivation for studying the
interaction of photons with the nucleus are many, ranging from the 
fundamental nuclear structure studies to studying 
the inner working of a nuclear reactor and the inner processes in a
star. Concerning nuclear reactors, photonuclear reactions contribute
to the general performance of the reactor and their accurate
understanding is a necessary part of any reactor core simulation. At
the same time, the issues of nuclear astrophysics involve many
reactions relating to photons, since, in a star, they are
ubiquitous. In fact, photonuclear reactions are
crucial steps in many of the nucleosynthesis processes generating the
observed abundances of elements. Thus, many
photonuclear reaction experiments are aimed at and motivated by
astrophysical references to nucleosynthesis, see
\Ref{Webb:2005bm,Meyer:1992zz,Lambert1992,Rauscher2001,Arnould20031,Hayakawa:2004nw,Utsunomiya:2005xc}
and references there-in.

In addition, the photonuclear reactions have been frequently applied
in photo-activation studies pertaining to various fields. This combination
of basic research aimed at fundamental physics and applied research
makes the photonuclear reaction
quite interesting and facilitates active research in several fields. This activity is also
on the increase as the availability of radiation sources increases
and their cost decreases. For the study presented in this paper, the
availability of a good radiation source was the main impetus for
embarking on this research activity.

Photo-activation experiments involving linear accelerators have been
performed at several institutions around the world
\Ref{Weller:2009zza,paa}, most notably 
at specialized laboratories such as S-DALINAC at TU
Darmstadt \Ref{Mohr1999480,Sonnabend20116} and ELBE in
Forschungszentrum Dresden in Rossendorf \Ref{Schwengner2005211}.
However, a study in \Ref{Mohr:2007jm} has shown how a non-specialized instrumentation such as a
clinical linac can also be employed for such experiments. This idea is a departure
from conventional approaches in which linear accelerators used for photonuclear experiments
have been designed and commissioned with sole use in nuclear physics experiments in mind.

The concept of using an ''off-the-shelf'' linac, originally designed for a different
purpose, is a novel idea. Although this idea is an attractive one, the 
original study presented in \Ref{Mohr:2007jm} has not been followed up on 
and its impact has been very limited. In light of availability of 
clinical linacs - in Turkey alone there are over 200 such devices - 
it is clearly a wasted opportunity not to use them for research. 
Especially important is the potential
contribution such devices can have on progress of experimental 
nuclear physics in developing countries. In developing countries
access to specifically designed linacs is practically non-existent, 
whereas access to clinical linacs is often readily available. 
Thus either trying to obtain one such linac, after its duty cycle in
medicine is over, or working closely with the local hospital, to utilize
the linac still in us in its off-hours, presents a
way how these device can help developing countries to contribute to the
global nuclear physics knowledge development. It is important 
to note further that there has been a great expansion of available 
radiation sources in the form of clinical linacs not just in
developing countries, but across the world.

The main aim of this work is to follow up on the original 
work of \Ref{Mohr:2007jm} and revive this line of research.
We will show how a repurposed clinical linac can produce
solid nuclear physics experiments, and that it is possible even without
a sophisticated specifically designed linac to obtain good quality data. What we intend to demonstrate is that 
nuclear physics experiments with a clinical linac are a viable concept
and one that can be of value to the global science community.

For this purpose it is essential to choose an appropriate 
example where the opportunity for contribution can be 
clearly demonstrated. A kind of niche where limited budget
experiments can contribute to the global knowledge. 
One such opportunity presents itself in cases where 
the measurements of basic nuclear quantities, such as
transition energies and half-lives, have not been performed
for several decades even in the case of elements near to the 
valley of stability. Hopefully by presenting an appealing and 
convincing example the viability of the concept will be clearly
demonstrated contributing to its popularity. 

In order to best illustrate the potential of clinical linacs in experimental 
nuclear physics, we have chosen to focus on zinc
isotopes to study the transition energies and half-lives of
isotopes created by photonuclear reactions. 
The choice of zinc as a target of bremsstrahlung photons produced by
the clinical linac is motivated by several reasons. Mainly, we are
interested in studying intermediate mass nuclei, especially
those isotopes on the proton-rich side of the nuclear chart.
In addition, the data for the $^{63}$Zn isotope, the main focus of this study,
and its $\beta$-decay product $^{63}$Cu are nearly 40 years old, see
\Ref{Klaasse1974}. This is the second part of the
motivation for this study,  \emph{i.e.} to revisit some of the
experimental results obtained quite some time ago and not
investigated again. The goal behind this choice is to
improve the data accuracy thus
illustrating what this and similar studies can accomplish.
This work aims at demonstrating both the feasibility of
photonuclear experiments with a clinical linac as well as to point out that
there are isotopes, close to the valley of stability, whose
reexamination is worthwhile. In the initial study presented 
in \Ref{Mohr:2007jm} the second part of this motivation 
was not explored in detail. In addition, we would like to 
note that we have already had
reasonable success with this approach in our preliminary study published in
\Ref{firstpaper}.

The paper is organized as follows; in \Sec{sec:lin} we give the
properties of the clinical linac used in this study. Detector setup
and the experimental procedure used are presented in \Sec{sec:exp}.
The explanation of the data analysis and the presentation of the
results are provided in \Sec{sec:res} and finally, we present our
summary and conclusions in \Sec{sec:conc}.

\section{Properties of the clinical linac}
\label{sec:lin}

\begin{figure}[!h]
  \begin{center}
    \centerline{\hbox{
    \includegraphics[width=0.45\textwidth]{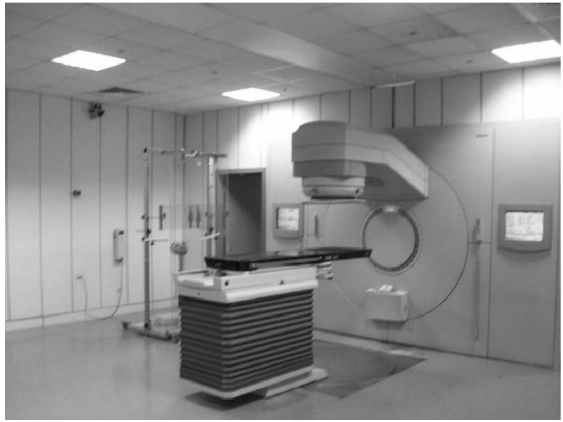}
}}
 \caption{Philips SLI-25 clinical linear electron accelerator of Elekta TM Synergy TM.}
    \label{fig:linac}
  \end{center}
\end{figure}

The photonuclear reactions by their very nature require a source of
photons, energetic enough to excite the target nuclei. In our
experiment, the photon source was a clinical electron linac. The
linac was re-commissioned from its usual role in medical physics for
the use in the study of photonuclear reactions. It should be noted
that a pioneering study by \Ref{Mohr:2007jm} demonstrated that these
clinical linacs (cLINAC) have properties appropriate for use in
photonuclear reactions. In fact, it was demonstrated that they even
compare well with the linacs made specifically for physics
experiments. And while our study has been conducted with a re-commissioned
linac the study in \Ref{Mohr:2007jm} was done while the linac was still
in medical use. While the later arrangement is possible with some coordination
and planing the former arrangement gives more freedom for conduction experiments.

In our setup, we have used a cLINAC SLI-25 manufactured by Philips
Medical Systems (currently part of Elekta TM Synergy TM), 
as a bremsstrahlung photon source, shown in \Fig{fig:linac}. The accelerator's
technical documentation can be found in \Ref{Electa}. The cLINAC
primary electron beam is generated by an electron gun with an energy
of about 50 keV. The electron gun in SLI-25 is a diode design with a
400 Hz pulse repetition frequency. After injection into the linac's
copper cavity, the electrons are accelerated by a radio-frequency
wave with 3 GHz (2856 MHz), S-band. The copper cavity is a
traveling wave design, where the power is injected at the beginning
of the accelerator structure. Even though the linac is designed to
operate at energies up to 25 MeV, the power is provided by the
magnetron instead of the klystron, which is more common for such
energies. The nominal power supplied by the magnetron is 2.5 MW at 4
MeV (low energies) and 5 MW at 25 MeV (high energies). As is usual,
all steering and focusing of the beam is achieved by standard
magnetic and electrostatic devices. 

\begin{figure}[!h]
  \begin{center}
    \centerline{\hbox{\hspace{0. cm}
    \includegraphics[width=0.47\textwidth]{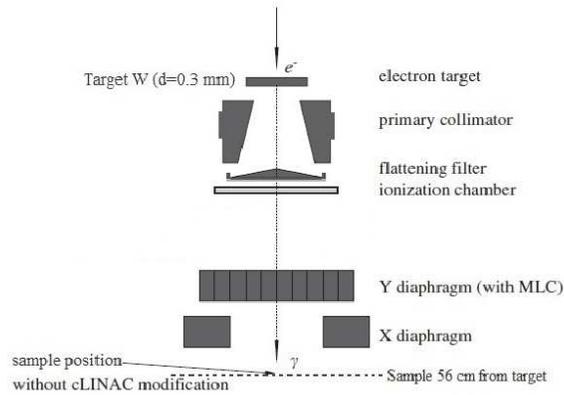}
}}
 \caption{Schematic view of the cLINAC.}
    \label{fig:scheme}
  \end{center}
\end{figure}

\begin{figure}[!h]
  \begin{center}
    \centerline{\hbox{\hspace{0. cm}
    \includegraphics[width=0.47\textwidth]{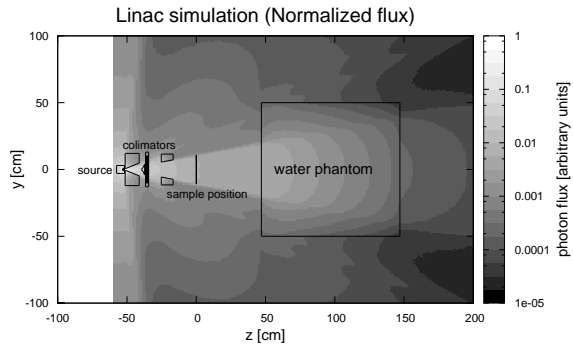}
}}
 \caption{Linac photon-flux distribution simulation normalized to the maximal flux.}
    \label{fig:sim}
  \end{center}
\end{figure}

After exciting the copper cavity, the beam is steered through
112.5$^\mathrm{o}$ slalom-type magnetic and electrostatic devices
before falling onto a high-$Z$ element target, in our case,
tungsten. The tungsten target is 0.3 mm thick and serves as an
electron stopper and a bremsstrahlung photon source. Because the
linac's original purpose was medical, \emph{i.e.} for patient treatment,
the bremsstrahlung photons are collimated and flattened with several
filters placed as shown in \Fig{fig:scheme}. The resulting photon beam 
is spatially uniform with no position dependence, \emph{i.e.} the beam 
has the same intensity at the center as it has at the edges of the 
field. To further illustrate this, we show in \Fig{fig:sim} a simulation 
of the photon flux created by the linac. As is evident in the figure
the photon flux is flat across the sample position at 56 cm.
For us, this is especially
convenient as it avoids the necessity for complicated arrangements
to create uniform sample irradiation. The focusing and collimation
is a standard feature of all cLINACs, as it is paramount to maintain
excellent spatial dose profile control when irradiating a patient.
In fact, it is common to require dose knowledge to better than 3\%
accuracy. The dose and the dose spatial distribution are measured
regularly as part of the standard linac performance monitoring.

\Fig{fig:sim} shows a simulation of the photon flux 
produced by the linac. The figure shows the photon flux relative to
the maximum flux in arbitrary units. It shows the location of the collimators
and the flattening filter as well as their influence on the flux.
Once the beam passes the flattening filter, which is of conic shape and much 
thicker in the center than at the edges, the beam loses any spatial dependence
in intensity and becomes flat across the width of the filter.
In addition, the collimators strongly reduce the beam intensity outside 
the focal area leading to a difference of several orders of magnitude 
in intensity inside and outside of the focal opening.
After the flattening filter and the collimators the beam remains
both flat and focused as it continues its trajectory.
Once it reaches the sample target
position at 56 cm, an area of approximately 20$\times$20~cm$^2$, it still has
a uniform beam profile. The scoring plane for the bremsstrahlung spectrum 
was placed at the surface of the water phantom at 100 cm downstream. At this 
distance the beam covers uniformly an area of 40$\times$40~cm$^2$. Note
that the surface of the water phantom was greater than 40$\times$40~cm$^2$.
Even as it travels through the phantom the beam still retains its
uniformity to a very good degree. Although not used in the present experiments
the collimators and the flattening filters can be used to create narrower 
or irregularly-shaped field shapes. 

\begin{figure}[!h]
  \begin{center}
    \centerline{\hbox{
    \includegraphics[height=0.48\textwidth,angle=270]{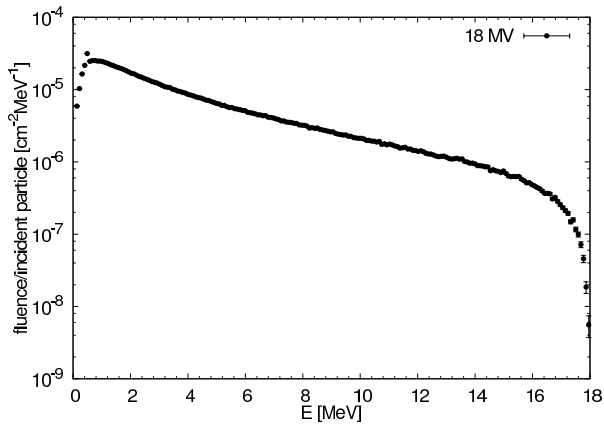}
}}
 \caption{EGS simulation of bremsstrahlung energy distribution from an electron
 beam accelerated over an 18 MV potential difference impacting
 on thick tungsten target in Philips SLI-25 clinical linac. The distribution shown is 
per incident electron.}
    \label{fig:beam18}
  \end{center}
\end{figure}

Under all of these conditions, the question of bremsstrahlung energy
distribution, which comes from the linac, becomes a quite
complicated one due to presence of all the filters and collimators.
A simulation of the photon energy distribution coming from the
SLI-25 was performed using the BEAMnrc package, \Ref{BEAMnrc}. The resulting
distribution is shown in \Fig{fig:beam18}. The distribution shown
was calculated with an electron beam accelerated over an 18 MV
potential difference impacting on a tungsten target of 0.3 mm thickness. As is
usual, the distribution was calculated per incoming electron. For
the estimation of the full flux, it is necessary to take into
account the number of incident electrons. In our case, this number
can be estimated from the dose delivered by the
beam, 5~Gy/min \Ref{Electa}, to be about $10^{11}~\mathrm{electron/s}$.
Combined with the simulation from BEAMnrc, this gives about $5\times
10^{5} photons/(\mathrm{MeVcm}^2\mathrm{s})$ at $\bar{E}=6$ MeV.

\section{Detector set-up and Experimental procedure}
\label{sec:exp}

In the experiment, one zinc sample was placed immediately outside
the cLINAC head at about 56 cm source to sample distance, see
\Fig{fig:scheme}. The sample in question was a zinc disk
weighing 5 g with 5 mm diameter and 1 mm thickness. 
The sample was irradiated for about 35 min. The time was optimized
such that a good count rate was achieved in the detector; the count
rate was low enough not to cause pile up in the detector and yet
high enough to observe several relatively weak peaks (intensity $\sim$ 0.05\%). The sample was
made primarily of zinc (67\% by mass fraction), but it had several
impurities, primarily magnesium, some aluminum, little iron and
trace amounts of other elements. However, in the previous tests, we
have already observed that the impurities did not show up in the
spectrum. In addition, since no special effort was made to
isotopically enrich the target, the presence of different zinc
isotopes can be considered as those of their natural abundances
($^{64}$Zn - 49.17\%, $^{66}$Zn - 27.17\%, $^{67}$Zn - 4.04\%,
$^{68}$Zn - 18.45\% and $^{70}$Zn - 0.61\%).

After the irradiation, the sample was transported to the Physics
Department of Akdeniz University, where the detector was located.
The detector used is a high-purity Germanium detector (HPGe). It is a
p-type, coaxial, electrically cooled HPGe detector, placed in a well-shielded
cavity. The shield is 10 cm thick lead with an inner surface covered
by a 2 mm copper foil to reduce the Pb X-rays generated in the Pb shielding. The
HPGe detector used is a gamma-ray spectrometer from AMETEK-ORTEC (GEM40P4-83)
with 40\% relative efficiency and resolution of 768~eV FWHM at 122~keV for
$^{57}$Co source and 1.85~keV FWHM at  1332~keV for $^{60}$Co source
\Ref{MAESTRO32}. It is connected to a set of Nuclear Instrumentation Modules
consisting of ORTEC preamplifier,
bias supply, spectroscopy amplifier, analog-to-digital converter and
a computer. Data acquisition was carried out with MAESTRO32
software. Through some trial and error in preliminary studies with
zinc, we came to the conclusion that the best setup of electronics
and detector for this experiment is the one with 16830 channels and
a channel step of about 0.18 keV/channel, giving about 3000 keV for
the last channel and the maximum transition energy which can be
observed.

The sample was placed in front of the detector about 10 min after the
irradiation and $\gamma$-ray counting continued for three days. During the counting, spectra
were automatically recorded at regular time intervals. Initially,
those time intervals were short $\sim$3 s, designed to follow the
short-lived isotopes, while the later ones became longer $\sim$20
min when focusing on longer-lived isotopes.

Immediately before counting, a set of calibration sources
were measured. For calibration, we used two sets of sources. The
point sources were supplied by the \c{C}ekmece Nuclear Research and
Training Center (IAEA 1364-43-2) and contained Co-60, Na-22, Mn-54,
Cd-109, Co-57, Cs-137 and Ba-133 isotopes. The second set is a soil
sample provided by the Turkish Atomic Energy Authority (TAEK). It
contained different natural radioactive isotopes
($^{40}$K,$^{226}$Ra and $^{232}$Th) with known activities. From
this sample, we used only the strongest peaks for calibration.

After the $\gamma$-ray measurement of the sample was performed, an equivalently long natural
background spectrum was recorded. Once the background spectrum was
recorded, the experiment was concluded with a second measurement of
the calibration sources. The aim of this second calibration-sources
measurement was to check the stability of the electronics and the
measurements. In this way, we were able to track any channel shift
during the measurement and also, through combining the before and
after calibration results, eliminate or at the very least reduce any
systematic errors coming from the channel shift.

The process of photo-activation applies to all the nuclei for which
the bremsstrahlung radiation exceeds the reaction threshold. In a
way, it is reasonable to assume that all the processes that are physically
possible are realized inside the sample. 
The neutron separation energies of the zinc isotopes are in the
range from 7 to 12 MeV and the proton separation energies are in the
range from 7.5 to 11.5 MeV. Given that the endpoint energy of our
beam was 18 MeV, it is expected that all of the stable zinc isotopes
will get activated. However, due to limitations of
the experimental setup, not all $\gamma$-decays from zinc isotopes were observable or have been
observed. The photonuclear reactions whose signatures we have
observed in the spectrum are:

\begin{align}
^{64}\mathrm{Zn}+\gamma &\rightarrow ^{63}\mathrm{Zn}\;\;+n, \label{eq:zn63_1} \\
^{66}\mathrm{Zn}+\gamma &\rightarrow ^{65}\mathrm{Zn}\;\;+n, \label{eq:zn65_1}\\
^{70}\mathrm{Zn}+\gamma &\rightarrow ^{69}\mathrm{Zn}^*+n, \label{eq:zn70_1}\\
^{68}\mathrm{Zn}+\gamma &\rightarrow ^{67}\mathrm{Cu}\;+p.
\label{eq:zn68_1}
\end{align}

The most important limitation for this reduced list is the half-life
of the created nuclei. It is because of this that we do not see
evidence of $^{67}$Zn since the half-life of $^{66}$Cu is only 5.12
min. At the same time the half-life criteria works in our favor and 
we are able to observe a long-lived (13.76 h) isomeric state of 
$^{69}\mathrm{Zn}$.

Our experimental setup was not designed for direct observation of
the above reactions, {\emph i.e. prompt gammas}, thus we have to rely on the radioactive decay of
the created nuclei and their end-stage $\gamma$-decays. The experiment was intended to
observe the half-life of the radioactive nuclei produced and the
transition energies of their daughter products, not to observe the
levels of stable nuclei which are used as the targets. In this
respect, we list the decay reactions which we studied

\begin{align}
^{63}\mathrm{Zn}\;  &\rightarrow ^{63}\mathrm{Cu}^*+e^+ +\nu, \label{eq:zn63_2}\\
^{65}\mathrm{Zn}\;  &\rightarrow ^{65}\mathrm{Cu}^*+e^+ +\nu, \label{eq:zn65_2} \\
^{69}\mathrm{Zn}^*  &\rightarrow ^{69}\mathrm{Zn}\;\; + \gamma, \label{eq:zn70_2}\\
^{67}\mathrm{Cu}\;  &\rightarrow ^{67}\mathrm{Zn}^*+e +\bar{\nu}.
\label{eq:zn68_2}
\end{align}

Here too, we have yet another limitation as to what we can observe. As
is noticeable the $\beta$-decay of $^{69}$Zn is missing from the above list
since it was not directly present in the observed gamma spectrum.
The observational limitation here is the low branching
ratio of the $^{69}$Zn $\beta$-decay into the excited states of $^{69}$Ga. Namely this
$\beta$-decay goes, almost exclusively (99.998\%), to the ground state of
$^{69}$Ga leaving no gamma transition to be observed by the detector.

In addition to the process of photonuclear reactions, explained above,
it should be noted that there exists a small ambiguity in the reactions for a
few of the cases presented. This ambiguity appears due to somewhat high
endpoint energy of the bremsstrahlung spectrum. As can be noticed,
at 18 MeV the photon energy is higher than the neutron separation energy for most elements.
As such, it is expected that the radiation source and the collimation
materials will become sources of neutrons. Therefore, some of the nuclei observed
could have been produced in neutron-capture reactions. However, this
source of neutrons as secondary particles is rather low and the processes
of photonuclear reactions presented above are dominant. Therefore,
for the results on which this work focuses, such small interferences are 
not of significance. Only for the cross-section determination can
such issues play a role.

These neutrons coming from the radiator and the collimation
materials in the head of the linac are of interest 
in medical therapy and can be measured.
In our case we have measured the slow neutron flux as being close to 
8000~$\text{neutrons}/(cm^2 s)$ at 100 cm distance from the radiator.
The measurement was performed in the center of the photon beam. However 
it is expected that the neutrons will have a similar presence even outside 
of the photon beam, since the radiator emits neutrons in all directions.
In medical therapy with photons this presents a 
problem since it causes a non-negligible dose outside the intended area.
Other treatment modalities with charged particle offer better dose profiles
and less neutron leakage \Ref{kaderka2012out,la2012out}.

\section{Data analysis and results}
\label{sec:res}

The data acquisition was performed with MAESTRO software;
however, due to its limitations peak analysis and energy calibration
were performed with other programs. Peak analysis was performed 
with RadWare code developed for analysis of gamma-ray
coincidence data,
by David Radford of the Physics Division at Oak Ridge National Laboratory \Ref{Radford1995297}, 
while energy calibration was performed in ROOT developed by CERN group  \Ref{Brun199781}.
The motivation for using two different
programs was to strike a balance between the desired accuracy and
time consumption for the data analysis. At the end, the data were
combined in a simple calculation sheet, giving the transition-energy
value and the associated error.

Thus, all of the recorded sample and background spectra as well as the calibration spectra taken before and
after the experiment were analyzed with the RadWare package. The strength
of the RadWare package and its suitability for the analysis of spectra recorded
with a HPGe detector lies in the fact that it fits a Gaussian, a skewed
Gaussian, and a smoothed step function to any number of chosen
peaks. In this way, we determine the centroids, areas and respective
statistical errors of all peaks found in a spectrum.

The first step of the analysis was performed on the calibration data
taken before and after the experiment. As is usual, the peak
position in terms of channel, \emph{i.e.} centroid, was obtained.
The two calibrations were then combined taking into account both the
errors obtained during the fit as well as the 
distance between them, \Ref{gun2008fundamentals}.
In this way, we obtained a unified determination of calibration-source
peak centroids. These were paired with the corresponding transition-energy data
taken from literature. The best example of this procedure is the
Co-60 with its two peaks at
1173.237$\pm$0.004 keV and 1332.501$\pm$0.005 keV. 
The centroids and energies are fitted with the aid of ROOT and
the energy calibration with accurate uncertainty for the fit parameters 
is obtained. In addition, the fitting subroutine provides the correlation 
between the fit parameters. We note that the ROOT fitting procedure
takes into account both the uncertainties in energy as well as 
the centroid according to the effective variance method.

At this point, the question arises as to which is the appropriate
fitting function for the energy calibration. Nominally, the
energy calibration should be linear, but in practice it may not be.
Thus, we tested linear, quadratic and cubic energy calibrations. 
In our case, the quality of the fit improved as we used higher order
polynomials. In fact, the $\chi^2/\mathrm{n.d.f.}$ for linear fit was
15.48, for quadratic 5.65 and cubic 1.41. Hence, we used the cubic 
fit for the energy calibration.

The propagation of error from the defining energy calibration
equation $E=\sum_{i=0}^3 a_i ch^i$ then takes all errors into account.
Error of the fit parameters $\sigma_{a_i}$, covariance $cov_{ij}$ or
correlation $cor_{ij}$ matrix and the errors of the centroid
determination itself $\sigma_{ch}$. The error formula  can then be
written as:
\begin{align}
\sigma_E^2&=\sum_{i=0}^3\left(\frac{\partial E}{\partial
a_i}\right)^2\sigma_{a_i}^2 +2\sum_{i=0}^3\sum_{j>i}^3 
\left(\frac{\partial E}{\partial a_i}\right)\left(\frac{\partial
E}{\partial a_j}\right)\mathrm{cov}_{ij}+
\left(\frac{\partial E}{\partial ch}\right)^2\sigma_{ch}^2\\
&=\sum_{i=0}^3\left(\frac{\partial E}{\partial
a_i}\right)^2\sigma_{a_i}^2+2\sum_{i=0}^3\sum_{j>i}^3 
\left(\frac{\partial E}{\partial a_i}\right)\left(\frac{\partial
E}{\partial a_j}\right)\mathrm{cor}_{ij}\sigma_i\sigma_j\nonumber\\ &+ \left(\frac{\partial
E}{\partial ch}\right)^2\sigma_{ch}^2. \label{eq:eq_calib}
\end{align}

With the calibration done, we proceed to determine the peak
position in the measured zinc sample spectrum. 
In addition to the sample spectrum, we also determined the 
recorded peak position in the background spectrum and
used it to discard the same peaks found in the sample spectrum. 
What remained was the centroid position information for the peaks, which
could be assigned to the sample. The energy and uncertainty 
were calculated according to \Eq{eq:eq_calib} and are listed in
\Tab{tab:res}.

\begin{figure*}[!ht]
  \begin{center}
    \centerline{\hbox{
    \includegraphics[height=1.\textwidth,angle=270]{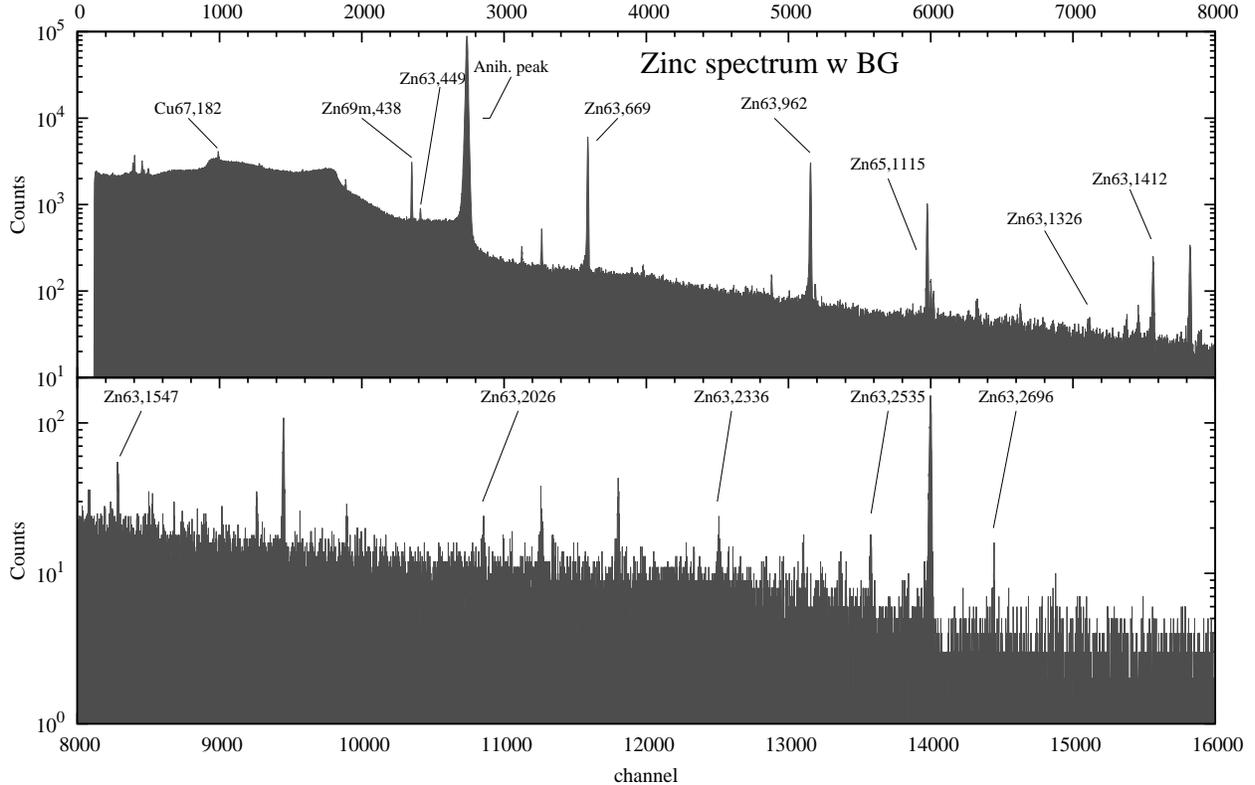}
}} \vspace{-0.7 cm}
 \caption{Zinc sample spectrum with no subtractions, for the irradiated sample after 3 days of counting.
 Assigned peaks are labeled with their energy in keV based on the energy calibration, and also with the decaying isotope.} %
    \label{fig:Zn_1}
  \end{center}
\end{figure*}

When we compared the energy results, from \Tab{tab:res},
with the literature values, it was
possible to assign these peaks to $\gamma$-ray transitions in specific isotopes. In \Fig{fig:Zn_1}, we show the irradiated-sample
spectrum without any background subtraction. The energy assignment of
peaks is shown in the figure. All the unassigned peaks are either
background peaks or sum and escape peaks. On the other hand, all the
assigned peaks can be connected with the nuclei produced in the photonuclear
reactions on zinc isotopes. Hence, it is safe to say that any elemental
pollution we might have had in the sample did not show up in the
spectrum in the form of an unidentifiable peak. In this way, we
verify that our assumption of de facto working with a pure zinc
sample was justified.

\begin{table*}
\begin{center}
\begin{tabular}{|c|c|c|c|c|c|c|c|}\hline
Ele.& $E_{N}$(keV)&$\sigma_{N}$ & $\bar{E}$(keV) & $\sigma_{E}$&$\Delta$=|$E_{N}$-$\bar{E}$| 
&D=$\sqrt(\sigma_{N}^2+\sigma_{E}^2)$&$\Delta$/D \\ \hline
\hline
$^{63}$Zn&449.93  &0.05 &  449.942  &0.017 &  0.012 & 0.05 & 0.24  \\
$^{63}$Zn&669.62  &0.05 &  669.69   &0.03  &  0.07  & 0.06 & 1.17  \\
$^{63}$Zn&742.25  &0.10 &  742.18   &0.03  &  0.07  & 0.10 & 0.70 \\
$^{63}$Zn&962.06  &0.04 &  962.09   &0.04  &  0.03  & 0.06 & 0.50  \\
$^{63}$Zn&1123.72 &0.07 &  1123.80  &0.07  &  0.08  & 0.10 & 0.80 \\
$^{63}$Zn&1327.03 &0.08 &  1327.11  &0.10  &  0.08  & 0.13 & 0.62  \\
$^{63}$Zn&1374.47 &0.13 &  1374.41  &0.08  &  0.06  & 0.15 & 0.40  \\
$^{63}$Zn&1392.55 &0.08 &  1392.56  &0.08  &  0.01  & 0.11 & 0.09  \\
$^{63}$Zn&1412.08 &0.05 &  1412.16  &0.11  &  0.08  & 0.12 & 0.67   \\
$^{63}$Zn&1547.04 &0.06 &  1547.16  &0.12  &  0.12  & 0.13 & 0.92  \\
$^{63}$Zn&2026.8  &0.3  &  2026.70  &0.18  &  0.10  & 0.4  & 0.25  \\
$^{63}$Zn&2336.5  &0.3  &  2336.4   &0.2   &  0.1   & 0.4  & 0.25  \\
$^{63}$Zn&2536.0  &0.3  &  2535.9   &0.3   &  0.1   & 0.4  & 0.25  \\
$^{63}$Zn&2696.6  &0.3  &  2696.5   &0.4   &  0.1   & 0.5  & 0.20  \\ \hline
$^{67}$Cu&184.577 &0.010&  184.63   &0.06  &  0.05  & 0.06 & 0.83  \\ \hline
$^{69}$Zn&438.634 &0.018&  438.601  &0.017 &  0.033 & 0.025& 1.32  \\ \hline
$^{65}$Zn&1115.539&0.002&  1115.51  &0.06  &  0.03  & 0.06 & 0.50  \\  \hline
\end{tabular}
\end{center}
\caption{Gamma-ray energies obtained in the present measurement by averaging the results of
before and after calibration measurements compared to values found in the literature (NUDAT).} \label{tab:res}
\end{table*}

Finally, the results obtained for the average energy and the combined variance 
are compared to literature values in \Tab{tab:res}. The literature results quoted are taken
from NUDAT and come from Nuclear data sheets publications, which for these 
elements are
\Ref{ERJUN2001147,Browne20102425,Junde2005159,Nesaraja20141}. In
turn Nuclear data sheets results are taken from individual publications:
\Ref{Klaasse1974}
for $^{63}$Zn decay, \Ref{PhysRevC.17.1822,Yalcin200563} for $^{67}$Cu decay, 
\Ref{Zoller196915,PhysRevC.1.744} for $^{69}$Zn decay, and
\Ref{Helmer200035} and others for $^{65}$Zn decay. 
In order to illustrate the quality of the agreement, 
we also show in \Tab{tab:res} how distant are our values from the literature ones. In addition,
we show the ratio of the distance and the combination of our and
literature uncertainties. It may be observed that
the agreement is good. All the results agree 
within 1.5$\sigma$ and most are within $\sigma$ or better. Even if 
the uncertainties are not combined, but simply the larger of the two is taken,
the agreement is still better than 2$\sigma$. 
At the same time, the results presented are of similar quality or better than
the literature ones. The only exception being $^{65}$Zn which is often 
used as a calibration standard and is measured to a very high accuracy.

In addition to the transition energies, the analysis performed
provided information about counts and their time evolution. This
information can be used to determine the half-life of the parent
nucleus as the decay of the daughter levels are in secular
equilibrium with the $\beta$-decay of the parent nucleus. Secular
equilibrium is a very good assumption here since the half-lives of
the states in question are orders of magnitude smaller than those of the
parents. The only exception is the $^{69}$Zn, which is an isomeric
transition and is thus observed directly.

Usually the measurement of the half-life has involved 
the measurement of the decay as function of time and the 
fit to the activity with the exponential decay curve ($A(t)=A_0exp(-\lambda t)$).
Also, a naive way would involve fits to the integral of the
activity which is directly represented by the counts.
However, fitting the peak in each consecutive step independently
and then obtaining the half-life from these data is not appropriate since
all the errors in each step are correlated, but the correlation
is unknown. A better approach is to integrate the activity
in equal-size time steps:
\begin{align}
C(T)&=\int_{T-\Delta T}^{T+\Delta T} A(t)dt =C_0e^{-\lambda T}\left(e^{\lambda \Delta T}-e^{-\lambda \Delta T}\right)
\label{eq:count}
\end{align}
where $C_0=A_0/\lambda$ and $T$ is counting time. So long as 
the $\Delta T$ is a constant the function only depends on $T$ 
exponentially just like activity.

In practical terms, this can be performed by taking independent spectra of length $\Delta T$
and restarting the count at the end of each time step. In this way, the counts obtained in two
consecutive spectra are no longer correlated and the errors of the counts obtained in this way
are independent of each other. Care should be taken to correct the counts obtained for dead time
since while the detector is busy the sample is still decaying. In our case, this was a minor correction
since only at the very beginning we had a small amount ($\sim 2$\%) of dead time.

In order to simplify the fit, logarithm of \Eq{eq:count} was used for fitting. In this
case, one has a simple linear fit from which the decay constant $\lambda$ is obtained. 
Once the value of the decay constant $\lambda$ has been obtained, 
it is a straightforward matter to calculate the half-life $T_{1/2}=ln~2/\lambda$.

\begin{figure}[!h]
  \begin{center}
    \centerline{\hbox{
    \includegraphics[height=0.46\textwidth,angle=270]{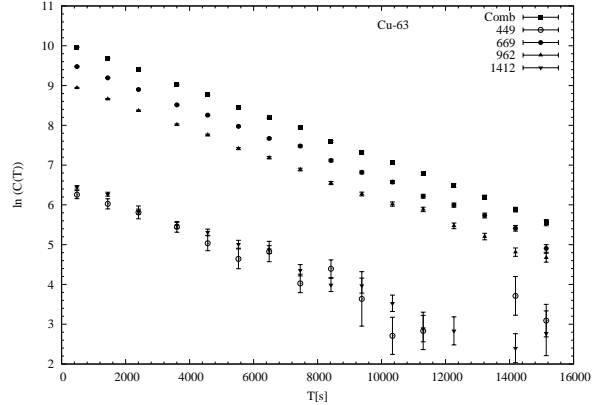}
}} 
 \caption{Logarithmic time evolution of $^{63}$Zn counts, as defined by \Eq{eq:count},
 for all transitions combined and separately for a few of the strongest ones.}
    \label{fig:Zn_hl}
  \end{center}
\end{figure}

Several examples of the counts dependence on time are shown in
\Fig{fig:Zn_hl}. Only the best are shown, \emph{i.e.} the strongest
peaks, as well as the dependence of the sum count of all peaks. As could have been
expected, a linear trend is quite evident and the error-bars are 
getting smaller for stronger peaks and earlier counting intervals.
The remaining peaks were too weak to be fitted, {\emph i.e.} 
the fitting procedure did not converge or gave very large uncertainties.

The summary of the fits to the function defined by \Eq{eq:count} 
and shown in \Fig{fig:Zn_hl} is given in \Tab{tab:half}. 
All of the errors listed in column 3 of \Tab{tab:half} are statistical only. 
As can be seen, the combined-counts fit offers the smallest error, but
the two strongest peaks 669 and 962 are also quite good.
When compared to the literature value of $38.47\pm
0.05$ min, the combined-counts fit is quite decent.  The
remaining fits scatter noticeably around these values. However,
all our values are systematically higher than the literature value.
This systematically higher result is likely the product of 
isotopic impurity of the sample, {\emph i.e.} the counts 
of a single isotope are distorted by the presence of other ones. 
Unfortunately in the current experimental setup we can not 
eliminate this systematic uncertainty.
Nevertheless, even with these pollution effects, the results 
are reasonable enough.

\begin{table}
\begin{center}
\begin{tabular}{|c|c|c|c|}\hline
Transition & $T_{1/2}$ [min] & $\sigma_{stat.}$&$\chi^2/\mathrm{n.d.f.}$   \\ \hline \hline 
Combined   &38.84 &0.15&0.782    \\ \hline 
449        &42.1  &3.3& 0.700       \\ \hline 
669        &38.70 &0.19&1.616    \\ \hline 
962        &39.19 &0.26&1.003    \\ \hline 
1412       &39.2 &1.2&0.977    \\ \hline 
\end{tabular}
\end{center}
\caption{$^{63}$Zn determined half-life from several transitions. These should be compared with NUDAT value of $38.47\pm 0.05$ min.}
\label{tab:half}
\end{table}

In addition to $^{63}$Zn, we also fitted the $\gamma$-decay of $^{69}$Zn isomeric
transition and obtained 13.76$\pm$0.18 h while the literature value
is 13.76$\pm$0.02 h in surprisingly good agreement with each other. Of the
other two observed $\beta$-decays $^{65}$Zn and $^{67}$Cu, we were not able
to fit their decay curves. The $^{65}$Zn was not fit due to unfavorable ratio
of its long half-life (244 d) and observation time (3 d). The
$^{67}$Cu was not fit due to the weakness of the signal. Like
in the case of $^{63}$Zn peaks, not listed in \Tab{tab:half}, it
had too big uncertainties to allow for a reasonable fit.

\section{Summary and conclusion}
\label{sec:conc}

In the experiment presented in this paper, we have investigated the
spectra and half-lives of zinc isotopes produced through photonuclear reactions.
The photonuclear reactions were induced by a
bremsstrahlung photon beam generated by a linac. The particularly
interesting point, in which we differed from previous such
experiments, is that we have used a clinical linac. Our aim was 
to demonstrate the potential usefulness and the power of such
machines, building on the pioneering work of \Ref{Mohr:2007jm}. 
Our work was motivated by the availability of clinical linacs. Their
presence is on the increase and after decommissioning from their use 
in medical treatments many are readily available to be used in research.
In addition even linac still in medical use can be utilized for 
nuclear physics research, with some planing and good coordination 
with the medical institution operating the linac.
Although this concept and the idea have been demonstrated in \Ref{Mohr:2007jm},
it has not created a trend of clinical-linac use in science.
With this paper, we wished to bring this idea back into focus.
Our aim was to show that a clinical linac can be a competitive tool
in modern nuclear physics, and that it can be especially useful for
laboratories such as ours, in the developing world, as an
easy and accessible way of performing experiments.

We used a beam of 18 MeV endpoint
energy, well above the proton and neutron separation
energies of all zinc isotopes, activating all of them. 
However, our experiment was not designed to
observe prompt gammas since it was offline. The elements, of which 
transitions from their nuclear levels were observed, are the decay products of
radioactive elements created by the photo-activation of zinc. These
transitions, as well as the half-lives of the parent nuclei, were
the intended goal and in this respect the experiment performed was
quite successful.

The experiment was separated into several parts,  the crucial ones
being the sample spectrum measurement, analysis and the
calibrations. On the calibration side, we have made substantial
efforts to understand the sources of errors and perform the
analysis as accurate and complete as possible. We have performed 
and combined two calibrations, before and after the sample 
measurement, in order to account for channel drift during the course
of the experiment. Such a procedure ensured that our data 
had a good robustness and was reliable.
We have paid close
attention to the fitted polynomial, choosing cubic as the best. 
For the sample spectrum, we paid close attention to any
background contribution especially when considering the area under
the peak, \emph{i.e.} the counts. The second part of the analysis, 
the study of parent nuclei half-lives, was devoted to this.

As for the results themselves, we have demonstrated quite
confidently that experiments such as ours improve the accuracy of 
transition energies. Indeed, we have obtained results, which are in
good agreement with the literature values, but notably often with reduced uncertainties. 
Our main contribution can be seen in the case of
$^{63}$Zn, whose levels we have consistently determined to an
accuracy level that is same or better than the one found in the literature.
Granted the measurement to which we compared
our data came from an experiment performed nearly 40 years
ago \Ref{Klaasse1974}, but this is exactly where we expect to 
improve on data from the literature. The results obtained and the
setup of the experiment make the study presented in this paper quite
interesting and a valuable contribution to the nuclear data set. In
fact, we firmly believe that such clinical-linac-based studies on proton-rich
nuclei can offer improvements of data in many intermediate-mass
nuclei. In accordance with this idea we are currently investigating 
the application of the our experimental setup to other nuclei. Currently
we have already performed several follow up measurements on nuclei 
such as Cl, Br, Sc, Ga, Sb and Pr and are in the process of analyzing the data.
The initial results show much promise.

On the half-life side, our data are not as nearly impressive as for
the transition energies. However, given the limitation of this
study, it is still a good check of consistency. Furthermore, with the
improvement of our experimental setup, primarily in target
preparation and choice of nuclei, we are confident that further
studies will offer better results on the half-lives as well. 
Obviously, increased experience in this kind of experiments will also
contribute to improvement of results in the future. In fact 
our follow up experiment on other nuclei indicate a good improvement
in the half-life determination.

\section{Acknowledgments}
\label{sec:acc}
We would like to thank the Akdeniz University Hospital for their
generous support. The project has also received administrative
support from the office of the Akdeniz University Rectorate. In
addition, the project and its various contributors have been
supported in part by TUBITAK-MFAG 114F220
and the 2216-Foreigner
Research program. The authors would also like to thank Dr. Deniz Savran
for very enlightening discussions regarding our data analysis in respect to
half-life determination.


%

\end{document}